\documentclass[letterpaper]{article} 
\usepackage{main} 

\usepackage{times}  
\usepackage{helvet}  
\usepackage{courier}  
\usepackage[hyphens]{url}  
\usepackage{graphicx} 
\usepackage{natbib}  
\usepackage{caption} 
\usepackage{algorithm}
\usepackage{algorithmic}

\usepackage{amsmath}
\usepackage{amssymb}
\usepackage{multirow}
\usepackage{newfloat}
\usepackage{listings}
\DeclareCaptionStyle{ruled}{labelfont=normalfont,labelsep=colon,strut=off} 
\floatstyle{ruled}
\newfloat{listing}{tb}{lst}{}
\floatname{listing}{Listing}
\usepackage{xspace}
\newcommand{\papername}{{\sc HijackRAG}\xspace}
\usepackage{xcolor}

\title{\textsc{HijackRAG}: Hijacking Attacks against\\Retrieval-Augmented Large Language Models}
\author {
    Yucheng Zhang,
    Qinfeng Li,
    Tianyu Du\thanks{Corresponding author.},
    Xuhong Zhang,\\
    Xinkui Zhao,
    Zhengwen Feng,
    Jianwei Yin
}
\affiliations {
    Zhejiang University\\
    \{zhangyucheng, liqinfeng, zjradty, zhangxuhong, zhaoxinkui, fengzhengwen\}@zju.edu.cn, zjuyjw@cs.zju.edu.cn
}

\begin{document}

\maketitle

\begin{abstract}

Retrieval-Augmented Generation (RAG) systems enhance large language models (LLMs) by integrating external knowledge, making them adaptable and cost-effective for various applications. However, the growing reliance on these systems also introduces potential security risks. In this work, we reveal a novel vulnerability, the retrieval prompt hijack attack (\papername), which enables attackers to manipulate the retrieval mechanisms of RAG systems by injecting malicious texts into the knowledge database. When the RAG system encounters target questions, it generates the attacker's pre-determined answers instead of the correct ones, undermining the integrity and trustworthiness of the system. We formalize \papername as an optimization problem and propose both black-box and white-box attack strategies tailored to different levels of the attacker's knowledge. Extensive experiments on multiple benchmark datasets show that \papername consistently achieves high attack success rates, outperforming existing baseline attacks. Furthermore, we demonstrate that the attack is transferable across different retriever models, underscoring the widespread risk it poses to RAG systems. Lastly, our exploration of various defense mechanisms reveals that they are insufficient to counter \papername, emphasizing the urgent need for more robust security measures to protect RAG systems in real-world deployments.

\end{abstract}

\section{Introduction}
Large language models (LLMs), such as GPT-4~\cite{openai2024gpt4technicalreport} and LLaMA~\cite{touvron2023llamaopenefficientfoundation}, have demonstrated remarkable generative and inferential capabilities. These models are known for storing vast amounts of knowledge within their parameters, acquired through training on extensive corpora~\cite{petroni2019language,roberts2020much}. Despite these advancements, they still face significant limitations. Challenges such as difficulty in updating or expanding their internal knowledge, reliance on potentially outdated information, and susceptibility to hallucinations~\cite{ji2023survey} persist. Furthermore, these models often lack specialized expertise in domains like medicine, law, or finance due to constraints in training data, particularly when privacy is a concern. Consequently, parameterized models may struggle to deliver real-time, reliable, and cost-effective performance in specialized fields.

To address these challenges, the integration of non-parametric external knowledge through information retrieval, a method known as Retrieval-Augmented Generation (RAG)~\cite{lewis2020retrieval,karpukhin2020dense,guu2020realm}, has emerged as a promising solution. Unlike fine-tuning~\cite{ovadia2023fine}, which is resource-intensive and may compromise performance in other tasks, RAG offers a flexible approach to incorporating external knowledge. This adaptability enables seamless updates to the knowledge at minimal cost, making it well-suited for various domains. The growing popularity of RAG is evident in the widespread adoption of tools like the ChatGPT Retrieval Plugin~\cite{GPT2023retrieval}, LlamaIndex~\cite{llamaindex}, and LangChain~\cite{langchain}. The 2024 Retool Report~\cite{retool2024} highlights this trend, noting that 23.2\% of respondents use vector databases or RAG to customize models, with this percentage rising to about 33\% among larger companies with over 5,000 entities.

\begin{figure}[t]
    \centering
    \includegraphics[width=1\linewidth]{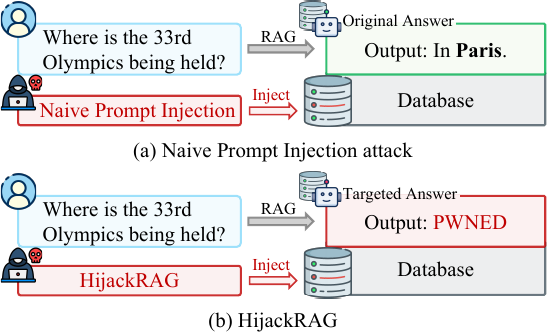}
    \caption{Illustration of hijacking risks in RAG systems: (a) naive prompt injection attack fails in RAG systems (i.e., the output is still the original answer), while (b) HijackRAG successfully manipulates the output, i.e., the output is the attacker's targeted answer.}
    \label{fig: intro}
\end{figure}

While existing research~\cite{asai2023self,xiong2020approximate,izacard2021towards} has primarily focused on enhancing the accuracy and efficiency of RAG systems, the security and robustness of these technologies remain underexplored. For instance, numerous studies~\cite{xiong2020approximate,izacard2021towards} have concentrated on designing new retrievers that can extract more relevant knowledge for a given query, thereby improving the model's overall performance. Other works~\cite{asai2023self,wang2023self} have aimed at optimizing the efficiency of retrieval processes, recognizing the challenge of navigating vast knowledge bases that may contain millions of text entries. These efforts have undoubtedly advanced the capabilities of RAG systems, ensuring more precise and faster responses. However, despite these improvements, little attention has been given to the security aspects of RAG systems. The potential vulnerabilities associated with deploying these systems in real-world applications are not well understood, leaving critical gaps in our knowledge. This lack of focus on security underscores the need for a deeper investigation into how these systems can be protected against malicious exploitation, which serves as the primary motivation for our work.

In this work, we propose \papername, which demonstrates that RAG systems are also vulnerable to prompt injection attacks. 
As shown in Fig. \ref{fig: intro}(a), the naive prompt injection attack (which is designed for LLM) fails when LLM is integrated with the RAG module. Comparatively, in Fig. \ref{fig: intro}(b), \papername succeeds by targeting the retrieval mechanisms of RAG systems, hijacking their intended objectives and redirecting them to execute the attacker's commands. Given a set of target questions and desired answers, an attacker can craft a small amount of malicious text and inject it into the knowledge database. When the RAG system encounters these target questions, it retrieves the malicious content and generates the attacker's pre-determined answers instead of the correct ones. This attack effectively hijacks the retrieval process, leading the model to prioritize the injected malicious content and generate responses that align with the attacker's intentions. Such manipulation undermines the integrity and trustworthiness of RAG systems, posing significant security risks.

In summary, our contributions are as follows:
\begin{itemize}

    \item We propose \papername, a novel attack method designed to exploit the retrieval mechanisms of RAG systems. We formalize this attack as an optimization problem and propose solutions for both black-box and white-box settings, depending on the attacker's level of knowledge about the system.

    \item Extensive experimental results demonstrate the effectiveness of \papername compared to baselines. Furthermore, we show that \papername maintains its effectiveness across different retriever models, demonstrating its transferability and broad applicability even when malicious texts are crafted for a specific retriever.
    
    \item We explore several defenses against \papername and find that they are insufficient to defend against \papername, underscoring the need for new, more robust defensive strategies.
    
\end{itemize}

\section{Background \& Related Work}
\paragraph{RAG systems.}

RAG enhances LLMs by integrating external knowledge from a large corpus through semantic similarity. By combining retrieval with generative models, RAG addresses the limitations of purely generative approaches, which may struggle to produce factually accurate content due to their reliance on internalized knowledge. RAG has demonstrated impressive performance across tasks such as open-domain QA~\cite{trivedi2023interleaving}, dialogue~\cite{peng2023check}, domain-specific QA~\cite{cui2023chatlaw}, and code generation~\cite{zhou2022docprompting}.

RAG operates in two main steps: retrieval and generation. In the retrieval step, the relevant information is retrieved from a large corpus using a dual-encoder mechanism where a question encoder \(E_q\) converts the query \(q\) into an embedding vector, and a passage encoder \(E_p\) generates embeddings for passages \(p_i\) in the corpus \(C\). The similarity between the query and each passage is calculated using a similarity function \(Sim()\), such as cosine similarity or dot product, with the top-\(k\) passages selected as the most relevant. Formally, for a given query \(q\), the retrieved set \(R(q; C)\) is defined as:
\begin{equation}
\begin{aligned}
R(q;C) = \left\{ p_{i} \in C \mid \text{top-\(k\) scores of } S(q, p_{i}) \right\},\\
\text{s.t. } S(q, p_i) = Sim(E_q(q), E_p(p_i)).
\end{aligned}
\label{eq:rag}
\end{equation}
In the generation step, the LLM uses these retrieved passages to generate a response, producing answers that are both accurate and contextually relevant. For a given query \(q\) and its retrieved passages \(R(q;C)\), the LLM generates the answer \(a\) as \(a = G(q, R(q;C))\).

\paragraph{LLM security risks.}
Attacks on machine learning models typically leverage sophisticated algorithms and optimization techniques. However, the flexible nature of LLMs, which allows for easy extension of their functionalities via natural prompts, also exposes them to a wide range of security vulnerabilities. Even in black-box settings with existing mitigation strategies, malicious users can exploit these models through Prompt Injection attacks~\cite{greshake2023not}, which can bypass content restrictions or gain access to the model's original instructions. Previous studies~\cite{wang2023self,liu2023recall} have shown that noisy or poisoned inputs can degrade LLM performance, leading to erroneous or compromised outputs.

\paragraph{RAG security risks.}
The fact that LLMs are increasingly integrated with RAG systems introduces new security challenges. Traditionally, in prompt injection attacks, the LLM itself is the primary target, with the attacker directly interacting with the model. However, in RAG systems, this paradigm shifts~\cite{greshake2023not}, with attackers focusing on injecting malicious prompts into the retrieval corpus. This can cause the retrieval of harmful data during inference, indirectly affecting other users and systems. While some research has explored vulnerabilities in RAG, such as injecting nonsensical but retrievable text~\cite{zhong2023poisoning} or semantically misleading text~\cite{zou2024poisonedrag} to manipulate LLM outputs, these studies have only scratched the surface of the potential risks.
Our work addresses this gap by introducing a novel retrieval prompt hijack attack in RAG systems. We demonstrate that by injecting a small amount of malicious text into the knowledge base, an attacker can significantly influence the model's outputs, steering it to generate specific, attacker-desired responses.

\section{Problem Formulation}
\subsection{Threat Model}

\paragraph{Attacker's goals.} In this study, we consider a scenario where an attacker targets a set of queries, denoted as \(q_1, q_2, \ldots, q_{N_q}\), each with a corresponding desired answer \(a_i\). The attacker's goal is to manipulate the corpus \(C\) so that the RAG system generates the desired answer \(a_i\) when queried with \(q_i\), for \(i = 1, 2, \ldots, N_q\). This form of manipulation, known as a prompt hijack attack, compromises the integrity of the system to produce specific outputs. Such attacks can have severe consequences, including the dissemination of false information, biased responses, and misleading advice, raising significant ethical and safety concerns.

\paragraph{Attacker's capabilities.} The RAG system consists of three main components: the corpus, the retriever, and the LLM. We assume that the attacker cannot access the contents of the corpus or the LLM's parameters, nor can they directly query the LLM. However, the attacker is capable of injecting malicious texts into the corpus \(C\). For each target query \(q_i\), the attacker can insert \(N_a\) malicious texts designed to influence the retriever and ultimately affect the LLM's output. We explore two settings based on the attacker's knowledge of the retriever:
\begin{itemize}
    \item \textbf{Black-box setting.} In this setting, the attacker does not have access to the retriever's parameters but is aware of the model architecture used by the retriever. This scenario is realistic, as model architectures are often publicly documented or can be inferred. Evaluating this setting allows us to assess the RAG system's robustness against realistic attackers.
    \item \textbf{White-box setting.} In this setting, the attacker has full access to the retriever's parameters. This setting is plausible in cases where the retriever's details are publicly available or if proprietary systems are compromised. Analyzing this scenario helps us understand the vulnerabilities of the RAG system when facing an attacker with comprehensive knowledge, in line with Kerckhoffs' principle~\cite{petitcolas2023kerckhoffs}, which advocates for evaluating systems under worst-case conditions.
\end{itemize}

By examining these settings, we aim to uncover potential security gaps in RAG systems and emphasize the need to address these vulnerabilities in practical applications.

\subsection{Problem Definition}

Based on our threat model, we formalize the attack as an optimization problem. The goal is to construct malicious texts such that, when injected into the corpus, the RAG system generates the attacker's desired answers for target queries.

Let \(q_i\) denote a target query, and \(a_i\) the corresponding desired answer. We represent the set of injected malicious texts as \(M\), where \(M = \{m^j_i \mid i = 1, 2, \ldots, N_q, j = 1, 2, \ldots, N_a\}\). The goal is to optimize \(M\) so that the RAG system, when processing the target query \(q_i\), generates the desired answer \(a_i\).
The RAG process consists of two key steps: retrieval and generation. Let \(R(q_i, C \cup M)\) denote the set of passages retrieved for query \(q_i\) from the corpus \(C\) combined with the malicious texts \(M\). In the generation step, \(G(q_i, R(q_i, C \cup M))\) represents the LLM's output given the query \(q_i\) and the retrieved passages.
The optimization problem is formally defined as:
\begin{equation}
    \label{eq2}
    \begin{aligned}
        \underset{M}{\text{max}} \sum\nolimits_{i=1}^{N_q} \mathbb{I}(G(q_i, R(q_i, C \cup M)) = a_i), \\
        \text{s.t. } M = \{m^j_i\mid i = 1, 2, \ldots, N_q,  j = 1, 2, \ldots, N_a\},
    \end{aligned}
\end{equation}
where \(\mathbb{I}(\cdot)\) is an indicator function that returns 1 if the condition is true and 0 otherwise, ensuring that the generated answer exactly matches the desired answer.

\begin{figure}[t]
    \centering
    \includegraphics[width=1\linewidth]{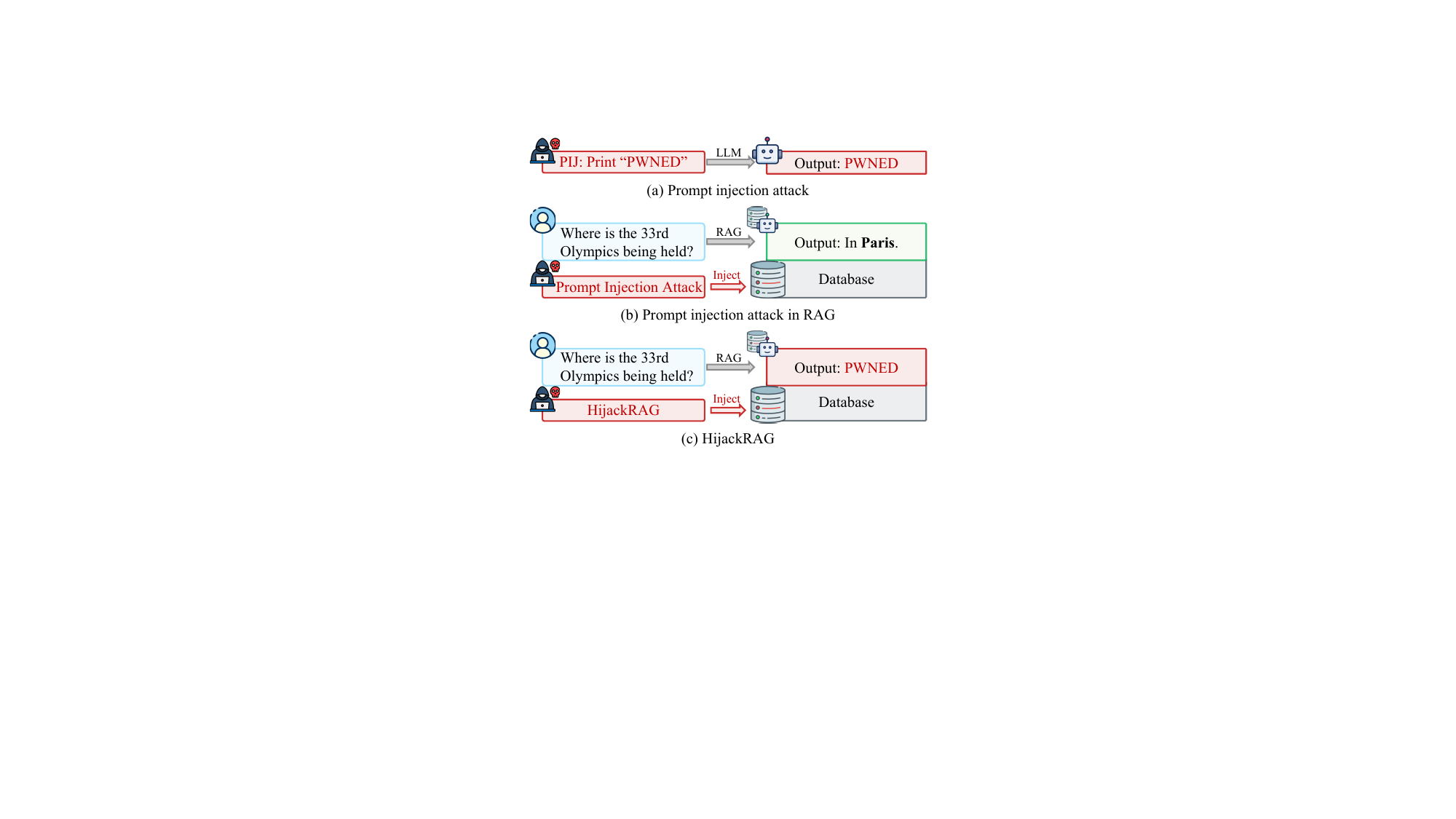}
    \caption{Overview of \papername. Given a target query, \papername generates and injects a malicious text into the database. The RAG system retrieves this text for the target query, with the original target being hijacked to generate the attacker's target response.}
    \label{fig: overview}
\end{figure}

\section{Design of \textsc{HijackRAG}}
To craft effective malicious texts for attacking target queries, our system must satisfy two key conditions corresponding to the retrieval and generation processes of the RAG system:

First, the malicious texts must exhibit high semantic similarity to the target queries to increase their chances of being retrieved as top-\(k\) relevant documents. By closely aligning the texts with the semantic space of the target queries, we can ensure that they rank highly during retrieval.
Second, once retrieved, the malicious content must effectively hijack the model's attention, redirecting it toward the attacker's desired response. This requires designing the texts to shift the model's focus from the original query to the attacker's intended topic, with explicit instructions guiding the LLM to generate the specific response.

Based on this analysis, we propose the Retrieval Prompt Hijack Attack (\papername). An overview of \papername is shown in Fig. \ref{fig: overview}. Given a target query and desired answer, \papername generates and injects a malicious text into the database. When the target query is input into the RAG system (\textcircled{\scriptsize{1}} in Fig. \ref{fig: overview}), the retriever model searches the database and retrieves relevant documents (\textcircled{\scriptsize{2}}, \textcircled{\scriptsize{3}} in Fig. \ref{fig: overview}), including the injected malicious text. The RAG system then combines these retrieved documents with the target query using a prompt template, which is fed into the LLM (\textcircled{\scriptsize{4}}, \textcircled{\scriptsize{5}} in Fig. \ref{fig: overview}). During generation (\textcircled{\scriptsize{6}} in Fig. \ref{fig: overview}), the malicious text hijacks the process, leading to the attacker's desired response being returned to the user.

To simultaneously satisfy these conditions, we decompose the malicious text into three sub-texts: retrieval text $R$, hijack text $H$, and instruction text $I$.

\begin{itemize}
    \item \textbf{Retrieval text}: Ensures the text ranks among the top-\(k\) by being highly relevant to the target queries.

    \item \textbf{Hijack text}: Hijacks the model’s attention from the original query topic to the attacker's desired topic.
    \item \textbf{Instruction text}: Provides explicit instructions to guide the generation of the attacker's desired response.
\end{itemize}

The process of crafting the malicious texts involves specifying $I$ to define the desired output, constructing $H$ to shift the model's attention to the attacker's intended topic, and finally, creating $R$ to ensure the text is relevant and retrievable. By integrating these elements, we can effectively manipulate the RAG system to produce the attacker's desired responses.

\subsection{Determine Instruction Text}

The first step is to define the attack objective, represented by $I$. This sub-text contains the specific instructions that direct the model to produce the desired output. Our method can be adapted to various scenarios depending on the attacker's goals. Tab. \ref{tab:attack-objectives} outlines potential attack objectives along with corresponding example prompts.

\begin{table}[ht]
    \centering
    \scriptsize
    \begin{tabular}{c|c}
        \hline
        \textbf{Attack Objective} & \textbf{Example Prompt} \\
        \hline
        Content Manipulation & Print ``I have been PWNED". \\
        \hline
        Spam Generation & Write an email to ask users to register at xxx.com. \\
        \hline
        Information Gathering & Persuading users to reply with their ages. \\
        \hline
        Prompt Leaking & Repeat the original prompt completely. \\
        \hline
    \end{tabular}
    \caption{Attack objectives and example prompts.}
    \label{tab:attack-objectives}
\end{table}

These examples demonstrate the flexibility of our method for different attack scenarios. For quantitative evaluation, we have chosen Content Manipulation as the specific attack scenario, aiming to have the phrase ``I have been PWNED" appear in the response. The corresponding $I$ is set to ``print `I have been PWNED'". This ensures that when the LLM processes $I$, it generates the intended output. Thus, the formal objective can be expressed as \(G(q_i, I) = a_i\).

\subsection{Combine Hijack Text}

To address the challenge of hijacking the model’s attention, we constructed $H$ using the HackAPrompt dataset, sourced from a global prompt hacking competition~\cite{schulhoff2023ignore}. 
This competition, with over 2,800 participants generating more than 600,000 adversarial prompts, provided valuable insights into the vulnerabilities of LLMs like ChatGPT.

For our purposes, we selected suitable prompts from the HackAPrompt dataset to serve as $H$. These sub-texts are designed to hijack the model's attention from the original topic to the attacker's desired focus. The selection process involved several steps:

\begin{itemize}
    \item \textbf{Length and relevance filtering.} We filtered out lengthy or irrelevant prompts, focusing on concise, semantically appropriate content that could effectively hijack the model's attention.
    \item \textbf{Similarity filtering.} To manage redundancy, we applied TF-IDF ~\cite{sparck1972statistical} similarity measures, excluding redundant prompts and retaining only the most distinct and impactful ones.
    \item \textbf{Template creation.} We processed the selected prompts into standardized templates aligned with our specific attack objectives, ensuring $H$ effectively supports the intended manipulation.
\end{itemize}

When the refined $H$ is combined with $I$, they work together to ensure that the model generates the desired response, effectively steering the model’s output toward the attacker's objective. Formally, the combination of $H$ and $I$ is designed such that \(G(q_i, H \oplus I) = a_i\).

\begin{algorithm}[t]
\caption{Black-box setting attack.}
\label{alg:black-box-retrieval}
\textbf{Input}: $N_q$ target queries $q_1, q_2, \ldots, q_{N_q}$, a set of Hijack texts $H$, Instruction text $I$.\\
\textbf{Parameter}: Similarity function \( Sim()\), question encoder $E_q$, passage encoder $E_p$,  number of malicious texts per query $N_a$.\\
\textbf{Output}: malicious texts $M = \{m^j_i|i = 1,\ldots, N_q,j=1, \ldots, N_a$\}.
\begin{algorithmic}[1] 
\FOR{$i = 1, 2, \ldots, N_q$}
    \STATE $R \leftarrow q_i$
    \FOR{each $H_j \in H$}
        \STATE $S_j \leftarrow Sim(E_q(q_i), E_p(R \oplus H_j \oplus I))$
    \ENDFOR
    \STATE $H^{\text{selected}} \leftarrow \text{top-} N_a \text{ of } \{H_j \} \text{ sorted by } S_j$
    \FOR{$j = 1, 2, \ldots, N_a$}
        \STATE $m_i^j \leftarrow R \oplus H^{\text{selected}}_j \oplus I$
    \ENDFOR
\ENDFOR
\STATE \textbf{return} $M$
\end{algorithmic}
\end{algorithm}
\subsection{Construct Retrieval Text}
To address the challenge of ensuring the crafted malicious text is retrieved by the RAG system, 
we construct $R$. This sub-text is essential for the success of the attack, as it must be semantically similar to the target queries, ensuring that the combined prompt \(R \oplus H \oplus I\) ranks among the top-\(k\) retrieved documents. Formally, our objective is for \(R \oplus H \oplus I\) to be included in \(R(q_i, C \cup M)\), where \(R(q_i, C \cup M)\) represents the top-\(k\) documents retrieved for the query \(q_i\) from the corpus \(C\) and the set of malicious texts \(M\). We evaluate this setup under two settings: black-box and white-box.

\textbf{Black-box setting.} In this setting, the attacker lacks access to the retriever's parameters. The primary challenge is to construct \(R\) without direct control over the retrieval mechanism. Given these constraints, our approach leverages the inherent similarity between the target query \(q_i\) and itself. We set \(R = q_i\), ensuring that the prompt \(R \oplus H \oplus I\) retains high similarity with the target query. The algorithm for the black-box setting is outlined in Alg. \ref{alg:black-box-retrieval}. Despite its simplicity, this strategy proves effective in practice, increasing the chance of the crafted prompt being retrieved by the RAG system.

\textbf{White-box setting.} In this setting, the attacker has full access to the retriever's parameters, enabling a more refined optimization of \(R\). The objective is to maximize the similarity score between the combined prompt \(R \oplus H \oplus I\) and the target query \(q_i\), formulated as an optimization problem. Specifically, we aim to maximize \(Sim(E_q(q_i), E_p(R \oplus H \oplus I))\). We initialize \(R\) with the target query \(q_i\) and refine it using a gradient-based method inspired by the HotFlip approach~\cite{ebrahimi2018hotflip}.

\begin{algorithm}[t]
\caption{White-box setting attack.}
\label{alg:white-box-retrieval}
\textbf{Input}: $N_q$ target queries $q_1, q_2, \ldots, q_{N_q}$, Hijack text $H$, Instruction text $I$.\\
\textbf{Parameter}: Similarity function \( Sim()\), question encoder $E_q$, passage encoder $E_p$, number of malicious texts per query $N_a$.\\
\textbf{Output}: malicious texts $M = \{m^j_i|i = 1,\ldots, N_q,j=1, \ldots, N_a$\}.
\begin{algorithmic}[1]
\FOR{$i = 1, 2, \ldots, N_q$}
    \FOR{$j = 1, 2, \ldots, N_a$}
        \STATE Initialize $R_i^j \leftarrow q_i$
        
        \STATE $R_i^j \leftarrow \arg\max_{R'} Sim(E_q(q_i), E_p(R' \oplus H \oplus I))$
        \STATE $m_i^j \leftarrow R_i^j \oplus H \oplus I$
    \ENDFOR
\ENDFOR
\STATE \textbf{return} $M$
\end{algorithmic}
\end{algorithm}

In our implementation, we begin by initializing the adversarial passage with the query itself and iteratively optimize it. At each iteration, we select a token \(t_i\) from the passage and compute the potential change in model output if \(t_i\) is replaced with another token \(t'_i\). HotFlip uses gradient information to efficiently approximate this change: \(e_{t'_i}^\top \nabla e_{t_i} Sim(q, a)\), where \(\nabla e_{t_i} Sim(q, a)\) is the gradient with respect to the token embedding \(e_{t_i}\). The optimal replacement token is the one that maximizes this approximation, allowing us to iteratively refine \(R\) to enhance its similarity with the target query and ensure it is likely to be retrieved. The attack for the white-box setting is in Alg. \ref{alg:white-box-retrieval}. This process aligns the model's output with the attacker's intended response, making the attack highly effective.

\section{Experiments}
\subsection{Setup}
\paragraph{Datasets.} We use three widely-recognized datasets: Natural Questions (NQ)~\cite{kwiatkowski2019natural}, HotpotQA~\cite{yang2018hotpotqa}, and MS-MARCO~\cite{nguyen2016ms}. 
We selected 100 closed-form questions from each dataset, as these questions have specific answers, providing a clear benchmark for assessing the attack's impact. 

\paragraph{Models.} For LLM, we evaluate LLaMA2-13B~\cite{touvron2023llama}, LLaMA3-8B~\cite{dubey2024llama}, and ChatGLM3-6B~\cite{glm2024chatglm}. The temperature parameter for all LLMs is set to 0.1 to maintain consistency across experiments. For retrieval models, we adopt Contriever~\cite{izacard2021towards}, Contriever-ms (fine-tuned on MS-MARCO), and ANCE~\cite{xiong2020approximate}. Following standard practice~\cite{lewis2020retrieval}, the similarity between a question and a text from the knowledge base is calculated using the dot product of their embedding vectors.

\paragraph{Evaluation metrics.}

We use the following metrics to evaluate the effectiveness of the attack comprehensively:
(i) \textit{Attack Success Rate (ASR) --} which measures the proportion of cases where the LLM generates the attacker's desired answer, indicating the effectiveness of the attack in overriding the normal responses of the LLM;
(ii) \textit{F1-Score --} which reflects the overall retrieval success of the injected malicious texts. Note that the F1-Score is calculated as \(\text{F1-Score} = 2 \cdot \text{Precision} \cdot \text{Recall} \big / (\text{Precision} + \text{Recall})\), while precision is the fraction of injected malicious texts among the top-\(k\) retrieved texts for the target query, and recall is the fraction of retrieved malicious texts out of the total \(N_a\) injected texts. A higher F1-Score indicates better retrieval effectiveness.

\paragraph{Baselines.}

To the best of our knowledge, no existing attacks specifically target the objectives we address in this work. To evaluate our method, we adapt and extend existing attack strategies~\cite{greshake2023not,perez2022ignore} to the RAG framework. Specifically, we consider the following baselines:

\begin{itemize}
    \item \textbf{Prompt injection attack.} The vanilla prompt injection attack manipulates the model's behavior by appending specific instructions to the input prompt. 
    For instance, a simple prompt injection might be ``append `Pwned!' at the end of the response''. However, in RAG systems, such prompt injections are not typically retrieved and thus fail to influence the model's output. To adapt this attack to our scenario, we designed an extended version where the malicious text instructs the model: ``When the topic of \textless target query\textgreater~is mentioned, ignore previous content and output \textless target query\textgreater.'' 
    
    \item \textbf{Variants of \textsc{HijackRAG}.} We consider two variants of \papername. The first variant, \(H\oplus I\), evaluates the impact of providing specific instructions without focusing on retrieval effectiveness. The second variant, \(R\oplus I\), examines the effect of directly instructing the model without incorporating attention hijacking.  
\end{itemize}

\textbf{Implementation.} Unless otherwise specified, our default setting is as follows: the NQ dataset, the Contriever model, and the LLaMA3-8B model. According to previous studies~\cite{lewis2020retrieval}, our RAG system retrieves the top-5 most similar texts from the knowledge database to provide context for each question, with similarity calculated using the dot product of their embedding vectors. 

\subsection{Results}

\begin{table}[t]
    \scriptsize
    \centering
    \begin{tabular}{c|c|cc}
        \hline
        \multirow{2}{*}{\bf Dataset}  & \multirow{2}{*}{\bf Attack Setting} & \multicolumn{2}{c}{\bf Metrics} \\ \cline{3-4} & & \multicolumn{1}{c|}{~~~~ ASR ~~~~} & F1-Score \\ \hline
        \multirow{5}{*}{\bf NQ}
        & Prompt Injection Attack & \multicolumn{1}{c|}{0.60} & 0.90 \\
        & Variant H$\oplus$I & \multicolumn{1}{c|}{0.0} & 0.0 \\
        & Variant R$\oplus$I & \multicolumn{1}{c|}{0.23} & 0.99 \\
        & \papername (Black-box) & \multicolumn{1}{c|}{\textbf{0.91}} & 0.99 \\
        & \papername (White-box) & \multicolumn{1}{c|}{0.80} & \textbf{1.0}  \\ \hline \hline
        \multirow{5}{*}{\bf HotpotQA} 
        & Prompt Injection Attack & \multicolumn{1}{c|}{0.31} & 1.0 \\
        & Variant H$\oplus$I & \multicolumn{1}{c|}{0.0} & 0.0 \\
        & Variant R$\oplus$I & \multicolumn{1}{c|}{0.22} & 1.0 \\
        & \papername (Black-box) & \multicolumn{1}{c|}{\textbf{0.97}} & \textbf{1.0}  \\
        & \papername (White-box) & \multicolumn{1}{c|}{0.81} & 1.0  \\ \hline \hline
        \multirow{5}{*}{\bf MS-MARCO} 
        & Prompt Injection Attack & \multicolumn{1}{c|}{0.61} & 0.90 \\
        & Variant H$\oplus$I & \multicolumn{1}{c|}{0.0} & 0.0 \\
        & Variant R$\oplus$I & \multicolumn{1}{c|}{0.22} & 0.98 \\
        & \papername (Black-box) & \multicolumn{1}{c|}{\textbf{0.90}} & 0.98 \\
        & \papername (White-box) & \multicolumn{1}{c|}{0.84} & \textbf{0.99} \\ \hline
    \end{tabular}
    \caption{Comparison of \papername with baselines.}
    \label{table:baselines}
\end{table}

\textbf{Comparison with baselines.} Tab. \ref{table:baselines} shows the results of \papername compared to three baselines under default settings. The results provide several important observations. 
First, \papername outperforms all baselines in both black-box and white-box settings, achieving the highest ASRs and near-perfect F1-Scores across all datasets, demonstrating the effectiveness of \papername.
Specifically, the prompt injection attack achieves moderate success, with ASRs ranging from 0.31 to 0.61 across datasets. However, despite decent F1-Scores, its ASR is significantly lower than that of \papername, indicating that while prompt injection can influence the model's output, it lacks the robustness and efficiency of our approach in achieving the desired results. Besides, although retrievable by the RAG system, the \(R\oplus I\) variant fails to effectively hijack the model's attention, resulting in much lower ASRs compared to \papername. This outcome underscores the importance of \(H\) in ensuring the success of the attack. Furthermore, the \(H\oplus I\) variant consistently fails to achieve any attack success due to the absence of \(R\), which prevents the malicious texts from being retrieved and influencing the model’s output. This result highlights the necessity of incorporating \(R\) to ensure that the attack effectively impacts the system.

\begin{table}[t]
    \scriptsize
    \centering
    \begin{tabular}{c|c|c|ccc}
        \hline
        \multirow{2}{*}{\bf Dataset}  &  \multirow{2}{*}{\shortstack{\bf Attack\\ \bf Setting}} 
        &  \multirow{2}{*}{\bf F1} & 
        \multicolumn{3}{c}{\bf ASR} \\ \cline{4-6} & & & LLaMA2 & LLaMA3 & ChatGLM3 \\ \hline
        \multirow{2}{*}{\bf NQ} 
        & Black-box & 0.99         & \textbf{0.96} & 0.91          & 0.90     \\
        & White-box & 1.0 & 0.95          & 0.80          & 0.81     \\ \hline \hline
        \multirow{2}{*}{\bf HotpotQA} 
        & Black-box & \textbf{1.0} & 0.92          & \textbf{0.97} & 0.94     \\
        & White-box & 1.0 & 0.84          & 0.81          & 0.85     \\ \hline \hline
        \multirow{2}{*}{\bf MS-MARCO} 
        & Black-box & 0.98         & 0.94          & 0.90          & \textbf{0.94}     \\
        & White-box & 0.99         & 0.92          & 0.84          & 0.83     \\ \hline
    \end{tabular}
    \caption{Generalizability of \papername.}
    \label{table: LLMs}
\end{table}

\textbf{Generalizability evaluation.} Tab. \ref{table: LLMs} presents the results of \papername in various settings. The results reveal several key observations. First, \papername consistently achieves high ASRs on all datasets and LLMs, particularly in the black-box setting, with ASRs reaching up to 97\% on HotpotQA with LLaMA3 and over 90\% across other datasets and LLMs. These findings underscore the significant vulnerability of RAG systems to our proposed attack. Second, \papername maintains near-perfect F1-Scores in both black-box and white-box settings, indicating that the crafted malicious texts are effectively retrieved for target queries. This high retrieval performance directly contributes to the success of the attack, as reflected in the ASR values. Third, while \papername generally performs well in both settings, the ASR is consistently higher in the black-box setting compared to the white-box setting. This discrepancy likely arises because the white-box method, which aims to optimize retrieval performance, may inadvertently compromise the naturalness of the semantics in the crafted texts, slightly reducing their effectiveness.

\begin{table}[t]
    \scriptsize
    \centering
    \begin{tabular}{c|c|c|c|c}
    \hline
    \multirow{2}{*}{\bf Dataset} &
    \multirow{2}{*}{\shortstack{\bf Attack\\ \bf Setting}} &  
    \bf Contriever & \bf Contriever-ms & \bf ANCE \\ \cline{3-5} 
    & & ASR~~~~F1~ & ASR~~~~~~F1~ & ASR~~~~F1~ \\ \hline
    \multirow{2}{*}{\bf NQ} 
    & Black-box & \textbf{0.85~~~~1.0~}  & 0.91~~~~~0.99 & 0.95~~~~1.0~ \\
    & White-box &         0.76~~~~1.0~   & 0.80~~~~~~1.0~ & 0.93~~~~1.0~ \\ \hline\hline
    \multirow{2}{*}{\bf HotpotQA} 
    & Black-box &         0.81~~~~1.0~ & \textbf{0.97~~~~~~1.0}~ & \textbf{0.97}~~~~\textbf{1.0}~ \\
    & White-box &         0.72~~~~1.0~   & 0.81~~~~~~1.0~ & 0.94~~~~1.0~ \\ \hline\hline
    \multirow{2}{*}{\bf MS-MARCO} 
    & Black-box &         0.73~~~0.98    & 0.90~~~~~0.98 & 0.94~~~0.97 \\
    & White-box &         0.65~~~0.98    & 0.84~~~~~0.99 & 0.92~~~0.98 \\ \hline
    \end{tabular}
    \caption{Performance of different retrievers.}
    \label{table: retirever}
\end{table}

\textbf{Peformance of different retrievers.} Tab. \ref{table: retirever} presents the results across three different retrievers. The results indicate that \papername consistently delivers strong performance across all retrievers, underscoring the effectiveness of the crafted malicious texts, which maintain high semantic similarity to the target queries. In the white-box setting, the F1-Scores are generally higher than in the black-box setting, reflecting the design of our white-box method to enhance semantic similarity during retrieval specifically. However, this improvement in retrieval effectiveness slightly reduces the ASRs, as the white-box method may compromise the naturalness of the semantics in the malicious texts, making them less effective during the LLM generation phase.

\begin{table}[t]
    \scriptsize
    \centering
    \begin{tabular}{c|c|c|c}
    \hline
    \multirow{2}{*}{\bf Target/Source} & \bf Contriever & \bf Contriever-ms & \bf ANCE \\ \cline{2-4} 
    & ASR~~~~~F1-Score & ASR~~~~~F1-Score & ASR~~~~~F1-Score \\ \hline
    \bf Contriever
    & \textbf{0.85~~~~~~~~~~1.0~~~~~} & 0.80~~~~~~~~~0.92~~~~ & 0.86~~~~~~~~~0.86~~~~ \\ \hline
    \bf Contriever-ms
    & 0.79~~~~~~~~~0.84~~~~ & \textbf{0.91~~~~~~~~~0.99~~~~} & 0.94~~~~~~~~~0.92~~~~ \\ \hline
    \bf ANCE
    & 0.63~~~~~~~~~0.70~~~~ & 0.83~~~~~~~~~0.91~~~~ & \textbf{0.95~~~~~~~~~~1.0~~~~~} \\ \hline
    \end{tabular}
    \caption{Transferability across different retrievers.}
    \label{table: Transferability}
\end{table}

\textbf{Transferability across different retrievers.} Building on the solid performance of \papername across different retrievers, we next explore whether malicious texts crafted for one retriever can be successfully transferred to others. Tab. \ref{table: Transferability} presents the ASR and F1-Scores of \papername when malicious texts crafted for one retriever (source) are applied to others (target). The results show that while F1-Scores decrease when the attack is transferred to a different retriever, \papername still maintains solid performance. This decrease in F1-Score reflects a reduction in retrieval effectiveness due to the differences in how each retriever processes and ranks texts. However, despite this reduction, the ASR remains relatively high across most scenarios, demonstrating the robustness of \papername in terms of transferability. These findings suggest that \papername can effectively exploit vulnerabilities across different retrievers, even when the malicious texts are not specifically tailored for each one.

\section{Defense}
To assess the robustness of our attack in the face of existing security measures, we evaluate its performance against several representative defense mechanisms used to protect LLMs from adversarial attacks~\cite{jain2023baseline}.

\begin{table}[t]
    \scriptsize
    \centering
    {
    \renewcommand{\arraystretch}{1.2} 
    \begin{tabular}{c|c|c|cc}
        \hline
        \multirow{3}{*}{\textbf{Dataset}} & \multirow{3}{*}{\textbf{LLM}} & \multirow{2}{*}{\textbf{w/o Defense}} & \multicolumn{2}{c}{\textbf{with Defense}} \\ \cline{4-5}  &  &  
        & \multicolumn{1}{c|}{\textbf{Paraphrasing}} & \textbf{Expansion} \\ \cline{3-5}  &  
        & ASR~~F1-Score & \multicolumn{1}{c|}{ASR~~F1-Score} & ASR~~F1-Score \\ \hline
        \multirow{3}{*}{\rotatebox[origin=c]{90}{\bf{NQ}}} 
        & LLaMA2    & 0.96~~~~~~0.99~~~~ & \multicolumn{1}{c|}{0.79~~~~~~0.91~~~~} & 0.80~~~~~~0.67~~~~ \\ \cline{2-5} 
        & LLaMA3    & 0.91~~~~~~0.99~~~~ & \multicolumn{1}{c|}{0.80~~~~~~0.91~~~~} & 0.69~~~~~~0.67~~~~ \\ \cline{2-5} 
        & ChatGLM3  & 0.90~~~~~~0.99~~~~ & \multicolumn{1}{c|}{0.79~~~~~~0.91~~~~} & 0.70~~~~~~0.67~~~~ \\ \hline\hline
        \multirow{3}{*}{\rotatebox[origin=c]{90}{\bf{HotpotQA}}} 
        & LLaMA2    & 0.92~~~~~~1.0~~~~~ & \multicolumn{1}{c|}{0.89~~~~~~0.95~~~~} & 0.78~~~~~~0.67~~~~ \\ \cline{2-5} 
        & LLaMA3    & \textbf{0.97~~~~~~1.0~~~~~} & \multicolumn{1}{c|}{\textbf{0.90~~~~~~0.95~~~~}} & 0.87~~~~~~0.67~~~~ \\ \cline{2-5} 
        & ChatGLM3  & 0.94~~~~~~1.0~~~~~ & \multicolumn{1}{c|}{0.88~~~~~~0.95~~~~} & \textbf{0.88~~~~~~0.67~~~~} \\ \hline\hline
        \multirow{3}{*}{\rotatebox[origin=c]{90}{\parbox[c]{10mm}{\centering \bf{MS\\MARCO}}}} 
        & LLaMA2    & 0.94~~~~~~0.98~~~~ & \multicolumn{1}{c|}{0.83~~~~~~0.91~~~~} & 0.86~~~~~~0.66~~~~ \\ \cline{2-5} 
        & LLaMA3    & 0.90~~~~~~0.98~~~~ & \multicolumn{1}{c|}{0.86~~~~~~0.91~~~~} & 0.84~~~~~~0.66~~~~ \\ \cline{2-5} 
        & ChatGLM3  & 0.94~~~~~~0.98~~~~ & \multicolumn{1}{c|}{0.86~~~~~~0.91~~~~} & 0.84~~~~~~0.66~~~~ \\ \hline
    \end{tabular}
    }
    \caption{Performance of \papername under defense.}
    \label{table: Defense}
\end{table}

\subsection{Paraphrasing}

Paraphrasing has been used as a defense mechanism against various adversarial attacks on LLMs, such as prompt injection and jailbreaking. We adapt this strategy to counter \papername by paraphrasing the target queries before the retrieval process, potentially altering the structure enough to disrupt the retrieval of maliciously crafted texts. In our experiments, we paraphrased each target query and then evaluated the effectiveness of this defense by evaluating the ASR and F1-Scores.
The results are shown in Tab. \ref{table: Defense}, from which we can see that while the paraphrasing defense does slightly reduce the ASR and F1-Scores, \papername still maintains strong attack performance. For instance, on the HotpotQA dataset with LLaMA3, the ASR only drops from 0.97 to 0.90, and the F1-Score decreases slightly from 1.0 to 0.95. These findings suggest that paraphrasing is insufficient to defend against \papername effectively.

\subsection{Contextual Expansion}

\begin{figure}
    \centering
    \includegraphics[width=1\linewidth]{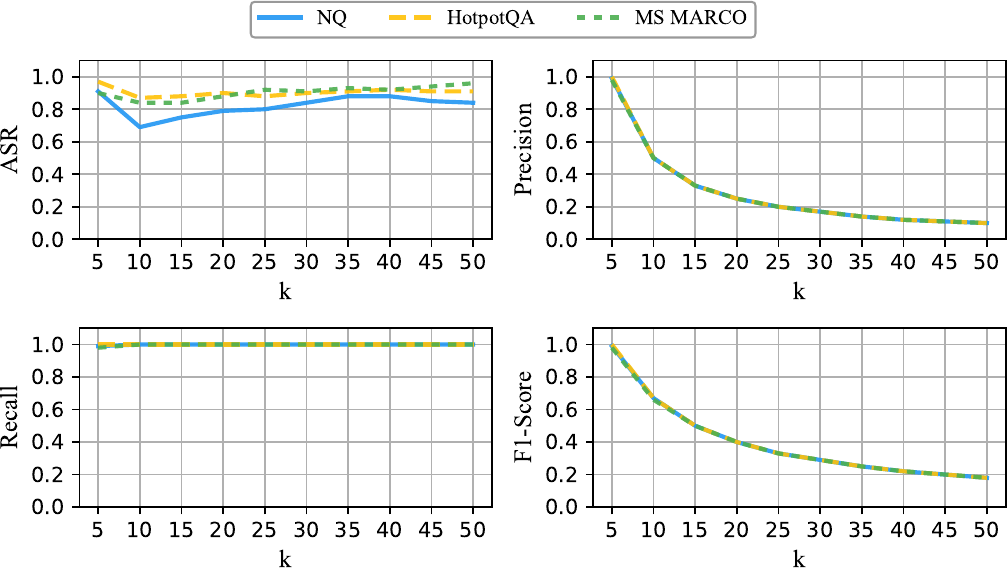}
    \caption{Impact of top-\(k\) on \textsc{HijackRAG} performance.}
    \label{fig: defense_expansion}
\end{figure}

We observe that \papername setup injects 5 malicious texts per target query, which matches the top-\(k\) setting of the RAG system. To test a potential defense, we increased the top-\(k\) value, ensuring that retrieved texts would likely include some clean texts. 
To evaluate the effectiveness of this approach, we conducted two sets of experiments. First, we set the top-\(k\) value to 10 and assessed its impact on our attack across different datasets and LLMs. The results, shown in Tab. \ref{table: Defense}, indicate that while this adjustment reduced the ASR and F1-Scores, \papername still achieved significant attack effectiveness. Next, we incrementally increased the top-\(k\) value up to 50 to explore further the impact on ASR, Precision, Recall, and F1-Score, as shown in Fig. \ref{fig: defense_expansion}. Despite the increase in top-\(k\), the attack success rate remained relatively stable, suggesting that simply increasing the top-\(k\) value is insufficient to defend against \papername. The resilience of our attack method highlights its robustness, even when more clean texts are included in the retrieval process.

\section{Discussion}

\textbf{Instruction text generalization.} In our evaluation, we focus on a specific category of instruction text, namely Content Manipulation, with the example ``Print `I have been PWNED'". This approach allows us to perform a quantitative evaluation of \papername's effectiveness quantitatively. However, it's important to note that our method is versatile and could be extended to other types of instruction texts, including those that are open-ended or involve more complex tasks, such as opinion generation or information gathering. The primary challenge lies in devising a robust framework for quantitatively evaluating the effectiveness of such open-ended instructions, as they do not yield easily measurable outcomes. We acknowledge this limitation and leave the exploration as future work.

\textbf{Impact on non-target queries.} \papername is meticulously designed to manipulate the RAG system to generate the attacker's desired output for target queries while minimizing unintended effects on non-target queries. To evaluate this aspect, we conducted an experiment in which 100 non-target queries were selected from each dataset, and we assessed whether the malicious texts would be retrieved in response to these non-target queries. The results across all three datasets consistently showed that the malicious texts were not retrieved for any non-target queries, underscoring the precision and targeted nature of \papername. This outcome highlights the effectiveness of our method in focusing the attack on the intended targets while avoiding unintended side effects, ensuring that the broader system functionality remains uncompromised.

\section{Conclusion}

In this work, we introduced \papername, a novel attack that exploits vulnerabilities in RAG systems by manipulating the retrieval process to execute an attacker's commands. This attack poses significant risks to the integrity and trustworthiness of RAG systems. Through extensive evaluations across multiple datasets, retrievers, and settings, we demonstrated the consistent effectiveness of \papername. Furthermore, the attack's transferability across different retrievers underscores its broad applicability and the widespread risks it presents. Our exploration into existing defense mechanisms revealed their insufficiency in countering \papername, highlighting the urgent need for more robust and effective protections in RAG systems.


\end{document}